\documentclass[preprint,1p,numbers,sort&compress,merge]{elsarticle}

\usepackage{lineno}
\usepackage{graphicx}
\usepackage{float}
\usepackage{bm}        % for math
\usepackage{amssymb}   % for math
\usepackage{slashed}   % for Dirac Slash
\usepackage{amsmath}   % for mutiline eqn
\usepackage{simplewick} % for contraction
\usepackage{verbatim}  % for multi-line comment
\usepackage{color} % for textcolor
\usepackage{xcolor} % for textcolor
\usepackage{appendix} % for appendix
\usepackage{subfig} % for sub fig
\usepackage{epstopdf}
\usepackage{hyperref}
 \usepackage{cprotect}
 \usepackage{verbatim}
\usepackage{listings}

\newcounter{YJC}

\graphicspath{{Figures/}}
\hypersetup{colorlinks = true,linkcolor = blue,anchorcolor =red,citecolor = blue,filecolor = red,urlcolor = red,
            pdfauthor=author}
            
\begin{document}
%\widetext

\title{MLAnalysis: An open-source program for high energy physics analyses}

\author{Yu-Chen Guo}
\ead{ycguo@lnnu.edu.cn}
\author{Fan Feng}
\author{An Di}
\author{Shi-Qi Lu}
\author{Ji-Chong Yang\corref{cor1}}
\ead{yangjichong@lnnu.edu.cn}
\cortext[cor1]{Corresponding author}

\address{Department of Physics, Liaoning Normal University, Dalian 116029, China}
\address{Center for Theoretical and Experimental High Energy Physics, Liaoning Normal University, Dalian 116029, China}

\begin{abstract}

We present a python-based program for phenomenological investigations in particle physics using machine learning algorithms, called \verb"MLAnalysis". 
The program is able to convert LHE and LHCO files generated by \verb"MadGraph5_aMC@NLO" into data sets for machine learning algorithms, which can analyze the information of the events. 
At present, it contains three machine learning (ML) algorithms: isolation forest (IF) algorithm, nested isolation forest (NIF) algorithm, kmeans anomaly detection (KMAD), and some basic functionality to analyze the kinematic features of a data set. 
Users can use this program to improve the efficiency of searching for new physics signals. 
 \\
Source code: \url{https://github.com/NBAlexis/MLAnalysis}\\
\textbf{Program summary}\\
Program Title: MLAnalysis\\
Programming language: Python3.8 and above\\
%External routines/libraries: NumPy, Matplotlib\\
Nature of problem:
With the continuous accumulation of experimental data, the research of high energy physics needs to process a large amount of data. ML methods can help us to improve the effect and efficiency of data analysis. Converting the data from experiments or Monte Carlo (MC) simulated events into data sets available for ML has become an important requirement. A program platform is needed for data preparation, as well as the application of various ML algorithms to improve the selection capability of target events and the efficiency of particle identification.
\\
Solution method: Supply an event analysis platform that supports ML approaches. The program is able to convert LHE and LHCO files into data sets that can be used for ML algorithms, and apply data preparation. 
In the data preparation step, the program transforms the raw data into a format that can be used to train and test machine learning algorithms, optimizes the adaptabilities and generalization capabilities of algorithms.
The program offers several algorithms, including IF, NIF, and KMAD, which provide NP model independent and standard model effective field theory operator independent methods to optimize event selection strategies. 
\end{abstract}


\maketitle

%%%%%%%%%%%%%%%%%%%%%%%%%%%%%%%%%%%%%
\section{\label{sec1} Introduction}
%%%%%%%%%%%%%%%%%%%%%%%%%%%%%%%%%%%%%

The search for new physics~(NP) beyond the Standard Model~(SM) is one of the most important tasks of high energy physics~(HEP).
In most cases, due to the good agreement between experimental measurements and the SM predictions, NP signals are expected to be rare events and their kinematic behaviors are different from that of the SM.
So it is necessary to optimize event selection strategies~(ESSs) with the help of the kinematic characteristics.
However, in some cases, it is difficult to efficiently suppress the backgrounds by kinematic cuts, so a better method for selecting signal events is needed.

Due to the fact that the NP signals are usually rare, a large number of events need to be analyzed both for experimental data and Monte Carlo (MC) simulation cases. 
Machine learning~(ML) methods can be helpful to improve the effect and efficiency of data analysis, and have been used in various aspects of HEP~\cite{mlreview,Baldi:2014kfa,Ren:2017ymm,Abdughani:2018wrw,DeSimone:2018efk,Ren:2019xhp,DAgnolo:2018cun,vanBeekveld:2020txa,ml1,CrispimRomao:2020ucc,Fol:2020tva,MdAli:2020yzb,Lv:2022pme}. 
With continuous improvements, ML has played an important role in particle identification~\cite{Baldi:2014kfa,Ren:2017ymm,Abdughani:2018wrw,Lv:2022pme}, searching for NP signals~\cite{Larkoski:2017jix,Guo:2018hbv,Abdughani:2019wuv,Li:2020fna,Kasieczka:2021xcg,Guo:2021jdn,Yang:2021kyy,Yang:2021ukg,Yang:2022fhw} and studying the polarization of final particles~\cite{Searcy:2015apa,Lee:2018xtt,Lee:2019nhm,Li:2021cbp}, etc.
The need for a program framework that can make ML algorithms more accessible is motivated.

In a typical ML algorithm, there are several main procedures involved, including: data preparation, model selection, training, evaluation, hyperparameter tuning, and deployment. 
In these steps, data preparation is a critical step in the ML process which is often overlooked but can have a significant impact on the performance of the model.
If we want to use ML algorithms to analyze experimental data or MC simulation events, we need to pre-process these data through data preparation. 
Data preparation is the transformation of raw data into a form that is more suitable for modeling. 
Data preparation involves data cleaning, data transformation, feature engineering. 
Data cleaning is to remove any missing values, correcting errors or inconsistencies. 
Data transforms are used to change the type of distribution of data variables and normalizing the data to fall within a certain range. 
Feature engineering is the process of selecting, transforming, and creating features that will be used to train the ML algorithms. 
Features are the variables that represent the input data and can be used to make predictions about the output. 
The goal of feature engineering is to create features that are relevant, informative, and non-redundant. 
This might involve selecting the most important features using techniques such as correlation analysis or feature importance scores, transforming the features using techniques such as scaling, normalization, or one-hot encoding, or creating new features that capture interactions or higher-order relationships between the original features. 
Data preparation is a critical step in a ML process, that can significantly impact the performance and accuracy of the final model. 

Recently, with the advancement of the study of the processes at the hadron collider with multi-neutrinos in the final states~\cite{Yang:2021ukg,Yang:2022fhw} and the optimization scheme of event identification~\cite{Guo:2018hbv,Yang:2021kyy}, we integrate the programs used in these studies into automatic program tools, called \verb"MLAnalysis".
%The readability and modifiability of this program with python code will be better than that of the package developed based on \verb"C++" framework.
The \verb"MLAnalysis" allows users to efficiently perform pre-defined and custom analyses of event files generated by the MC event generators such as \verb"MadGraph5_aMC@NLO".
The code has been tested in python 3.8 and above.

The importance of \verb"MLAnalysis" lies in its ability to help researchers unlock the value of the experimental or MC data. 
Many researchers collect large amounts of data in the course of their work, but may not have the expertise or resources to analyze it using ML. 
This program can help to bridge this gap by providing a user-friendly tool for data transformation and feature engineering that can help researchers to extract insights and knowledge from their data.
In addition to its data transformation and feature engineering features, this program also includes basic functionality for data processing and several ML algorithms.
This opens unlimited possibilities concerning the level of complexity which can be reached, being only limited by the programming skills and the originalities of the users.

The structure of the paper is as follows.
Section~\ref{sec2} overviews the structure and functions of \verb"MLAnalysis" package.
Then section \ref{sec3} introduces the ML algorithms provided.
Section~\ref{sec4} shows some general usages by examples.
Our summary can be found in section \ref{sec5}.

%%%%%%%%%%%%%%%%%%%%%%%%%%%%%%%%%%%%%
\section{\label{sec2} Program overview}

%%%%%%%%%%%%%%%%%%%%%%%%%%%%%%%%%%%%

The \verb"MLAnalysis" acts as a bridge between phenomenology studies and ML algorithms by transforming the data from experiments or MC into a format that can be recognized by ML algorithms, and also assists with feature engineering.
Some of the key features of the program include data cleaning and preprocessing, data transformation and normalization, feature selection and extraction, and data visualization. 
These features can help to automate and streamline the process of preparing the data for ML, saving time and reducing the risk of errors.
\verb"MLAnalysis" also includes basic functionality for data processing and several ML algorithms.
By including these additional features, \verb"MLAnalysis" provides a more complete suite of tools for working with phenomenological studies and ML algorithms. 

The purpose of this program is to make it easier for researchers to work with experimental data or the data from MC in the context of ML. 
By providing a user-friendly interface and a suite of tools for data transformation and feature engineering, \verb"MLAnalysis" can help to remove some of the barriers to entry for those new to the field of ML, and to make ML more accessible to a wider range of researchers in HEP.

\subsection{Data Structure}

The data structure of \verb"MLAnalysis" mainly includes five classes.
The \texttt{EventSet} is the uppermost container, which has only a list of objects with \texttt{EventSample} as the class.
The \texttt{EventSet} represents a set of collision events.
The \texttt{EventSample} is a class lists all particles of a collision event, this data structure contains a list of objects with \texttt{Particle} as the class.
The class \texttt{Particle} contains information about a particle, including particle type, four-momentum, mass, Particle Data Group identifier (PDG-id), etc.
The four-momentum is represented by the class \texttt{LorentzVector}.
The class of \texttt{Martix4x4} is used for operations on a \texttt{LorentzVector}, such as rotation and Lorentz boost.

\begin{table}[!htbp]
  \centering
  \cprotect\caption{The functions of the classes of \verb!MLAnalysis! and the types of input parameters of these functions.}\label{tab:structure}
   {\scalebox{.6}{ \begin{tabular}{|l|}
     \hline
\multicolumn{1}{|c|}{ EventSample}\\
     \hline
 particles: list \\
     \hline
AddParticle(particle: Particle)\\
DebugPrint(): str\\
     \hline
   \end{tabular}

\begin{tabular}{|l|}
     \hline
\multicolumn{1}{|c|}{EventSet}\\
     \hline
events : list \\
     \hline
AddEvent(event: EventSample)\\
AddEventSet(eventSet)\\
DebugPrint(i: int)\\
GetCopy()\\
GetEventCount(): int\\
     \hline
   \end{tabular}

\begin{tabular}{|l|}
     \hline
\multicolumn{1}{|c|}{LorentzVector}\\
     \hline
values: list \\
     \hline
Azimuth(): float\\
Et(): float\\
MakeWithRapidity(pseudoRapidity: float, azimuthal: float, pt: float, mass: float)\\
Mass(): float\\
P3d()\\
PseudoRapidity(): float\\
Theta()\\
V3d()\\
Y(): float\\
     \hline
   \end{tabular}
    }}\\

{\scalebox{.6}{
  \begin{tabular}{|l|}
     \hline
\multicolumn{1}{|c|}{Matrix4x4}\\
     \hline
values : NoneType, list \\
     \hline
MakeBoost(v3velocity)\\
MakeOne()\\
MakeRotation(degree: float, x: float, y: float, z: float)\\
MakeRotationFromTo(v3from, v3to)\\
MakeRotationFromToV4(v4from: LorentzVector, v4to: LorentzVector)\\
MakeZero()\\
MultiplyMatrix(otherMatrix)\\
MultiplyVector(vector: LorentzVector): LorentzVector\\
     \hline
   \end{tabular}

 \begin{tabular}{|l|}
     \hline
\multicolumn{1}{|c|}{Particle}\\
     \hline
PGDid : int\\
index : int\\
momentum\\
particleType\\
status \\
     \hline
DebugPrint(sep: str): str\\
SetLHCOOtherInfo(nTrack: float, bTag: float, hadEm: float)\\
     \hline
   \end{tabular}
      }}
  
\end{table}

In Table~\ref{tab:structure}, we show the functions of these classes and the types of input parameters of functions, where ``float'', ``str'' and ``int''  stand for the floating-point number, string, and integer, respectively.

\subsection{Import a data set}

Performing a phenomenological analysis on the results provided by MC generators or by experiments always starts with the reading of a set of event samples. 
Currently, the types of event files supported by \verb"MLAnalysis" include the Les Houches Event (LHE) files at the parton-level, and LHC Olympics data (LHCO) files at the reconstruction level. 
The files ``LesHouchesEvent.py'' and ``LHCOlympics.py'' in the ``Interfaces'' folder are used for importing. 
We classify all particle types into seven categories: ``jet'', ``electron'', ``muon'', ``tau'', ``photon'', ``intermediate'' and ``missing''.
It should be noted that, in an LHE file the ``missing'' refers to each neutrino, whereas in an LHCO file it is the sum of transverse momenta of all neutrinos.

\verb!def LoadLHCOlympics(fileName: str) -> EventSet:!

This function is fed with a full path to the LHCO file, and will return an object with \texttt{EventSet} as the class.

\verb!def SaveToLHCO(fileName: str, event: EventSet, realLHCO: bool = True):!

This function is fed with a full path to the LHCO file and an \texttt{EventSet}. 
The \texttt{EventSet} is read from an LHE or an LHCO file or built from the code. 
There might be incoming particles and intermediate particles.
Besides, there might be multiple neutrinos in an \texttt{EventSample}.
When ``realLHCO'' is turned on, only the outgoing particles are saved, and the neutrinos are combined into a missing transverse momentum.

\verb!def LoadLesHouchesEvent(fileName: str) -> EventSet:!

This function is fed with a full path to the LHE file, and will return an object with \texttt{EventSet} as the class.

\verb!def LoadLargeLesHouchesEvent(fileName: str, debugCount: bool) -> EventSet:!

Similar to \texttt{LoadLesHouchesEvent}, but reads an LHE file line after line.
When ``debugCount'' is turned on, a message will be printed after each \texttt{EventSample} is loaded.

The support for other file formats is on the way.

\subsection{Data cleaning and cuts}

Data cleaning is one of the steps in data preparation.
Besides, the ESS based on traditional kinematic cuts can be used to extract signals from backgrounds.
In both cases, a set of tools for cuts is necessary.
The files in the ``CutAndExport'' folder present the cut mechanism and some cuts for common usages.
To implement a cut, the ``CutEvents'' function can be used.

\verb!def CutEvents(eventSet: EventSet, cutFunction):!

The ``CutEvents'' function is fed with an \texttt{EventSet}, and an object ``cutFunction''.
It is required that the interface ``Cut'' is implemented for the ``cutFunction''.

\verb!def Cut(self, eventSample: EventSample) -> bool:!

The format of the interface ``Cut''. 
The ``Cut'' function should be fed with an \texttt{EventSample}.
Whether this \texttt{EventSample} should be cut off depends on the returned value of the ``Cut'' function.
If the ``Cut'' function returns ``True'', this event sample should be cut off, otherwise, should remain in the \texttt{EventSet}.

As an example, a typical class for ``cutFunction'' is presented as follows.
\begin{lstlisting}
class PhotonNumberCut:
    """
    If cutType = 0, cut all with photons > parameters[0]
    If cutType = 1, cut all with photons < parameters[0]
    If cutType = 2, cut all with photons not in parameters
    """

    def __init__(self, cutType: int, parameters):
        self.cutType = cutType
        self.parameters = parameters

    def Cut(self, eventSample: EventSample) -> bool:
        photonCount = 0
        for particle in eventSample.particles:
            if 0 == particle.particleType:
                photonCount += 1
        if 0 == self.cutType:
            return photonCount > self.parameters[0]
        if 1 == self.cutType:
            return photonCount < self.parameters[0]
        return photonCount not in self.parameters
\end{lstlisting}

The code defines a class named “PhotonNumberCut”, which has an initializer method that takes two arguments: ``cutType'' and ``parameters''. 

The ``cutType'' argument is an integer that determines how the event sample should be filtered based on the photon count, and ``parameters'' is a list of values that is used by ``cutType''.

The ``Cut'' method takes an ``EventSample'' object as an argument and returns a boolean value. 
This method counts the number of particles in the event sample that have a ``particleType'' of 0 (which represents photons) and compares it to the ``parameters'' based on the ``cutType''. 
If ``cutType'' is 0, it returns true if the photon count is greater than the first element of ``parameters''.
If ``cutType'' is 1, it returns true if the photon count is less than the first element of ``parameters''. 
If ``cutType'' is 2, it returns true if the photon count is not in the list of ``parameters''.

The ``CutAndExport'' folder also contains some unique cuts proposed in our anomalous quartic gauge coupling~(aQGC) and the neutral triple gauge coupling~(nTGC) studies~\cite{Guo:2019agy,Guo:2020lim,Guo:2021jdn,Fu:2021mub,Yang:2021ukg,Yang:2021kyy,Yang:2021pcf,Yang:2021kyy,Yang:2020rjt,Yang:2022fhw}. 
The ``Applications'' folder contains the complete project files for these studies.
 
Sometimes, the generic features might however not be sufficient according to the needs of the users.  
The projects in the ``Applications'' folder can provide references for users, and their analysis can be realized by using and modifying the relevant codes.  
Taking advantage of the Python interface, a user can define his own physics analysis in an efficient, flexible and straightforward way.

%%%%%%%%%%%%%%%%%%%%%%%%%%%%%%%%%%%%%
\section{\label{sec3}Methods of Machine Learning Algorithms}
%%%%%%%%%%%%%%%%%%%%%%%%%%%%%%%%%%%%%

The NP signals are generally rare and kinematically different compared with the SM background. 
Thus, the search for NP can be considered as anomaly detection (AD), then, ML algorithms can be used to find kinematically anomalous events.
\verb"MLAnalysis" contains ML algorithms for AD, and therefore can be used in the search of NP.

\subsection{\label{sec3.1} Isolation Forest Algorithm}

\begin{figure}[!htbp]
\centering{
\includegraphics[width=0.6\textwidth]{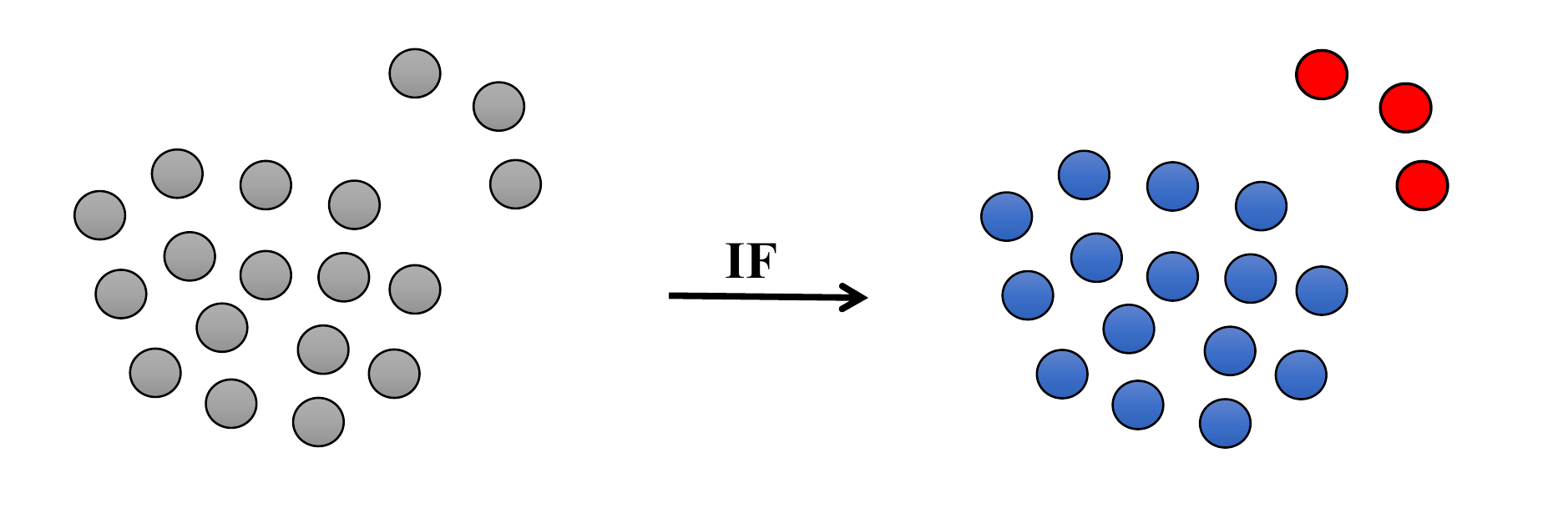}
\caption{\label{fig:IF} The IF algorithm can identify anomalous data. The signal points are the anomalous points colored in red, and the background points are colored in blue.}
}
\end{figure}

IF algorithm~\cite{4781136} is an unsupervised ML algorithm with linear complexity to deal with AD problems, which can effectively deal with large-scale multi-dimensional data.
It is good at finding the data which are ``few and different''.
Compared with normal samples, such anomalous samples are more easily isolated.
IF can be used to search for anomalous samples by constructing a binary tree structure, where the anomalous samples are closer to the root node compared with the normal samples~(see Fig.~\ref{fig:IF}).
Such binary trees are called Isolation Trees (iTrees).

\begin{figure}[!htbp]
\centering{
\includegraphics[width=0.4\textwidth]{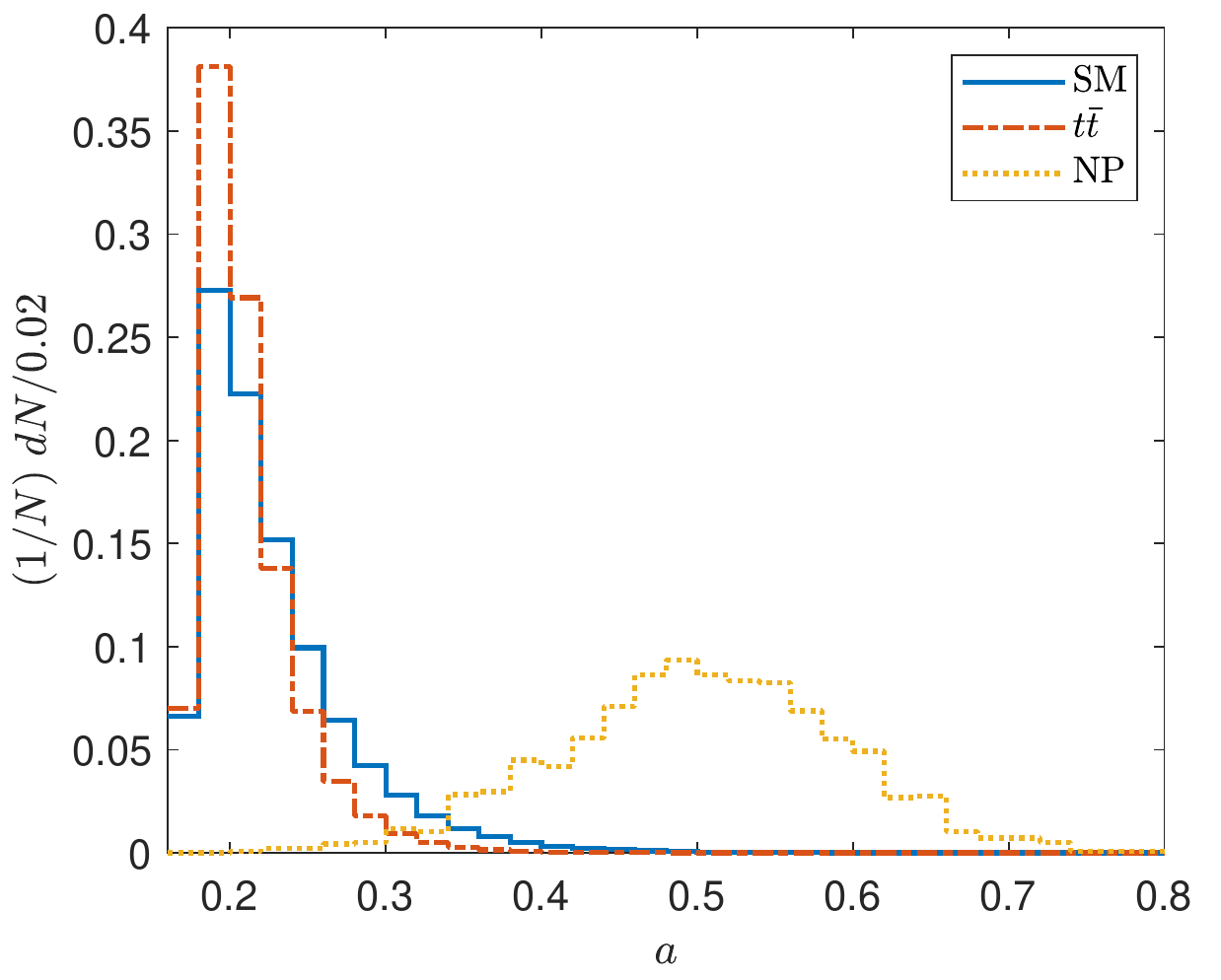}
\includegraphics[width=0.4\textwidth]{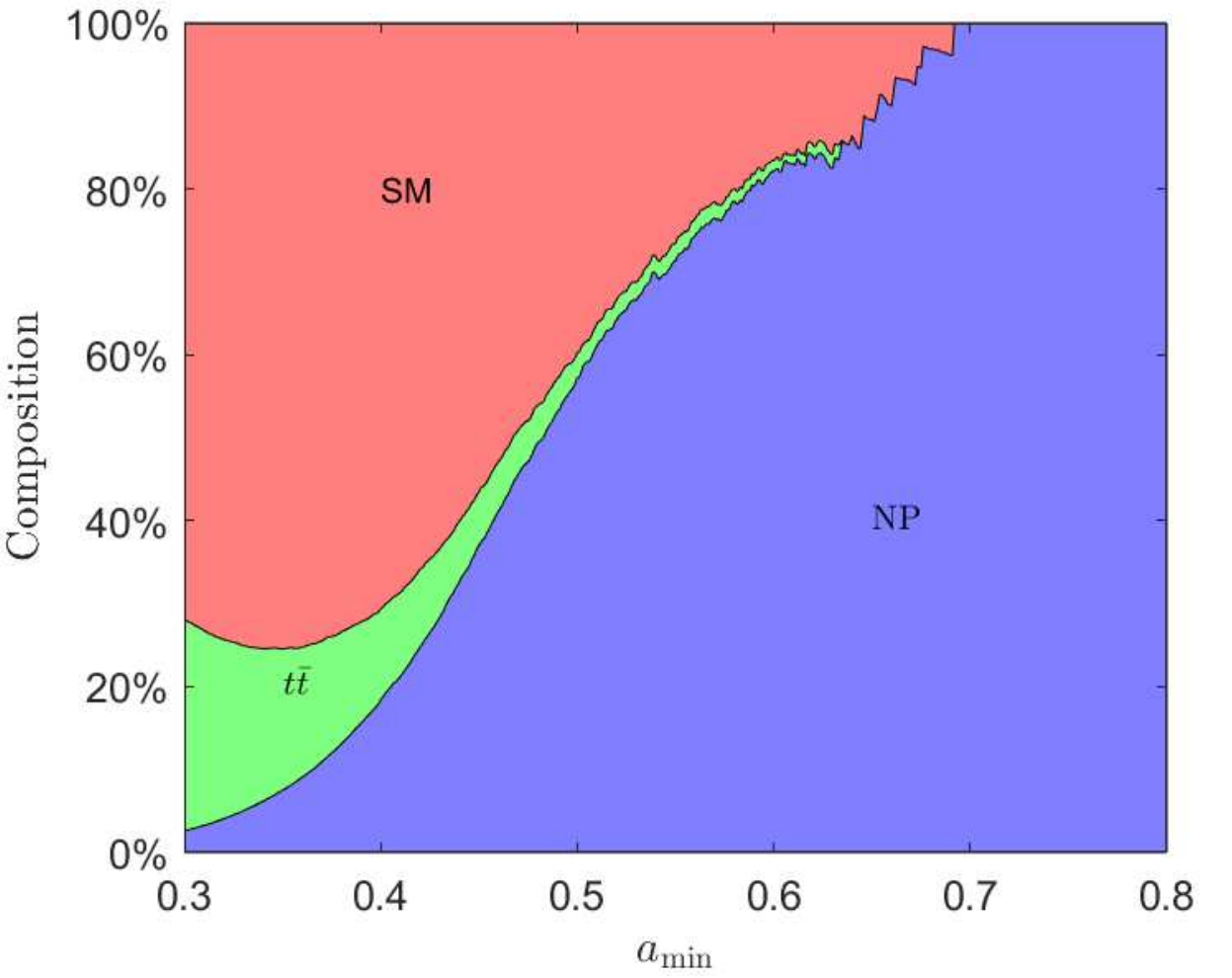}
\caption{\label{fig:dist}Normalized distributions of anomaly scores denoted as $a$ (left) and compositions of the selected events with $a>a_{\rm min}$, where $a_{\rm min}$ is a threshold to select the events (right).}}
\end{figure}

Here we use the vector boson scattering~(VBS) process $pp\to jj \ell^+\ell^-\nu\bar{\nu}$ with leptonic decay at the 13 TeV LHC as an example, which has been studied using IF~\cite{Guo:2021jdn}. 
The $t\bar{t}$ production with b-jet mistagged is also considered as the background.
The signal and background events are generated by \verb"MadGraph5_aMC@NLO"~\cite{madgraph}, with a parton shower by \verb"Pythia82"~\cite{pythia} and a detector simulation by \verb"Delphes"~\cite{delphes}.
IF algorithm is applied to calculate the anomaly scores for signal and background events.
It proves that the IF algorithm has the ability to distinguish between the SM signal and the background of the $t\bar{t}$ process,
as shown in Fig.~\ref{fig:dist}.

When using the IF algorithm, there is no need to know what kind of NP signal the data set contains, and there is no need to optimize the parameters according to the characteristics of the NP signal.
In other words, it is possible to select NP signals without a priori knowledge of the NP models.

\subsection{\label{sec3.2}Nested Isolation Forest Algorithm}

\begin{figure}[!htbp]
\centering{
\includegraphics[width=0.6\textwidth]{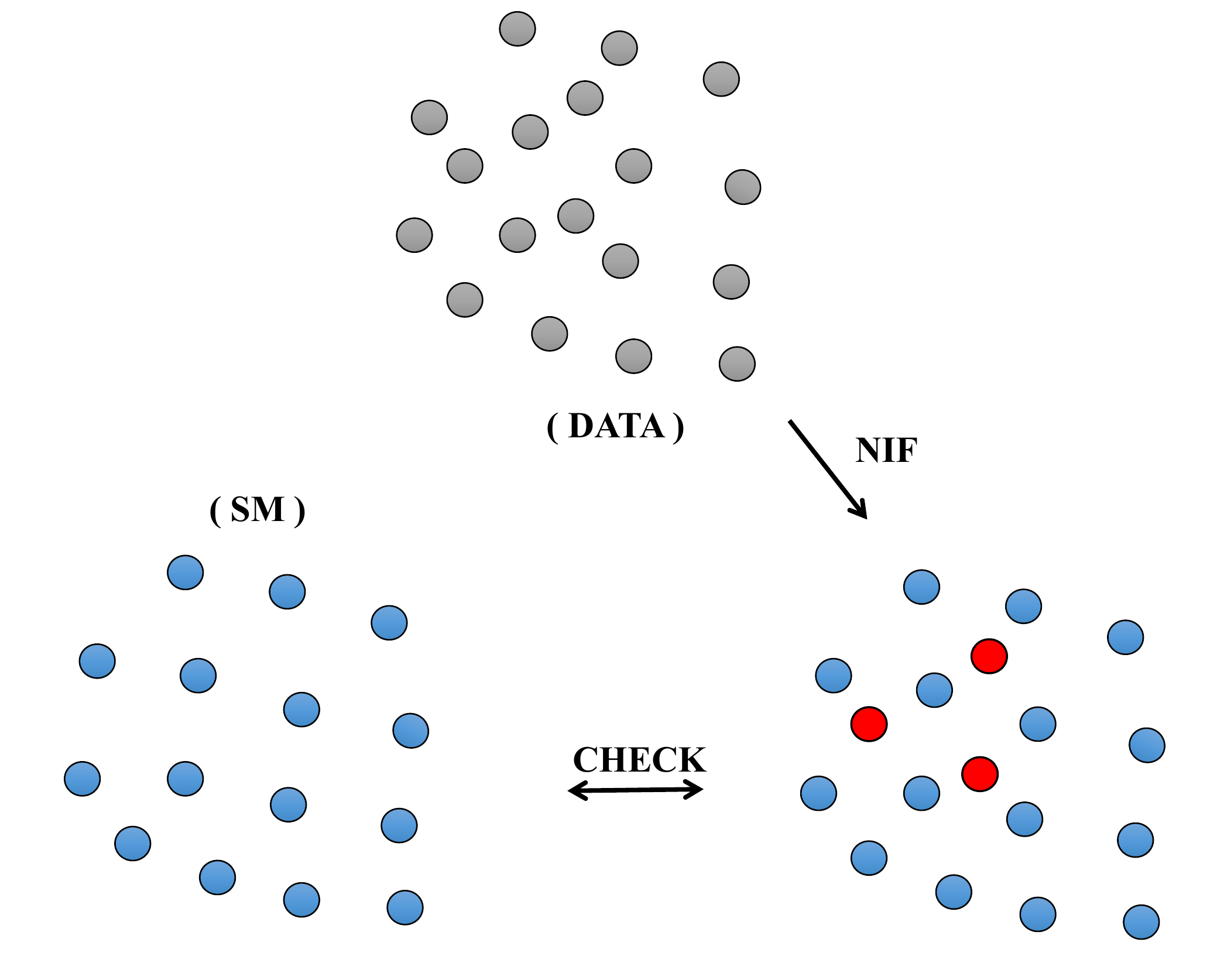}
\caption{\label{fig:problemofad} The distribution of signal points overlaps with the background, which is no longer an AD problem. By comparing the data distribution of the SM, NIF can select the points that change the distribution.}}
\end{figure}

Although the IF algorithm could select NP signals, AD-based algorithms were no longer applicable when the interference between NP and the SM is important.
To illustrate this, we use two dimensional points to describe this problem in Fig.~\ref{fig:problemofad}.
As can be seen from Fig.~\ref{fig:problemofad}, the signal in this case will increase the density of the data point distribution rather than been more easily isolated.
According to the IF, the anomaly scores would be higher in the area with low density. 
Thus, in the case of Fig.~\ref{fig:problemofad}, the anomaly scores of some points would be reduced when NP presents.
If the distribution of the anomaly scores for the SM events is taken as a reference benchmark, the variation of density in phase space can be measured by the change of anomaly scores. 
Then the existence of an NP signal can be detected in this way.

Based on this idea, we propose an unsupervised ML algorithm, which is called nested anomaly detection.
When an IF algorithm is nested, it is then nested IF~(NIF).
First, the MC simulation data set of the SM is used as the training data set, which is marked as $S_{\rm SM}$, and the anomaly scores $a_{\rm SM}$ of each event are obtained by the IF.
Then, the anomaly scores $a_{\rm data}$ for each event in the target data set $S_{\rm data}$ are obtained by the IF.
Finally, the closest events in the phase space between the target data set and the training data set are matched.
The change in anomaly score for each pair of events is $\Delta a^i =a^i_{\rm data}-a^i_{\rm SM}$.
Here the distance is defined as $d=\sqrt{\sum _{ij}(p_j^i-q_j^i)^2}$, where $p$ and $q$ are the 4-momenta of the particles in $S_{\rm data}$ and $S_{\rm SM}$, respectively, and $p_j^i$ and $q_j^i$ are the $i$-th component of the 4-momentum of the final-state particle $j$.
$\Delta a$ is the indicator used to detect the existence of NP signals. When the NP signal exists in the events, this indicator will be less than 0.
We can adjust the sensitivity of the NIF algorithm to the NP signal by setting the maximum value of $\Delta a$.

NIF algorithm not only inherits the advantages of IF, which is independent of NP models and SM effective field theory~(SMEFT) operators, but also solves the problem that AD can not be handled.
It has an intelligible operation mechanism and almost no adjustable parameters.
In addition, the NIF program framework of \verb"MLAnalysis" can be used not only for IF but also for any algorithm that can quantitatively measure the abnormal degree of each event, which results in a good generality.
The search of nTGCs in the process $e^+e^-\to Z\gamma$ at the $e^+e^-$ colliders is an example that the NIF algorithm works well~\cite{Yang:2021kyy}.

\subsection{\label{sec3.3} K-means anomaly detection method}

\begin{figure}[!htbp]
\centering{
\includegraphics[width=0.6\textwidth]{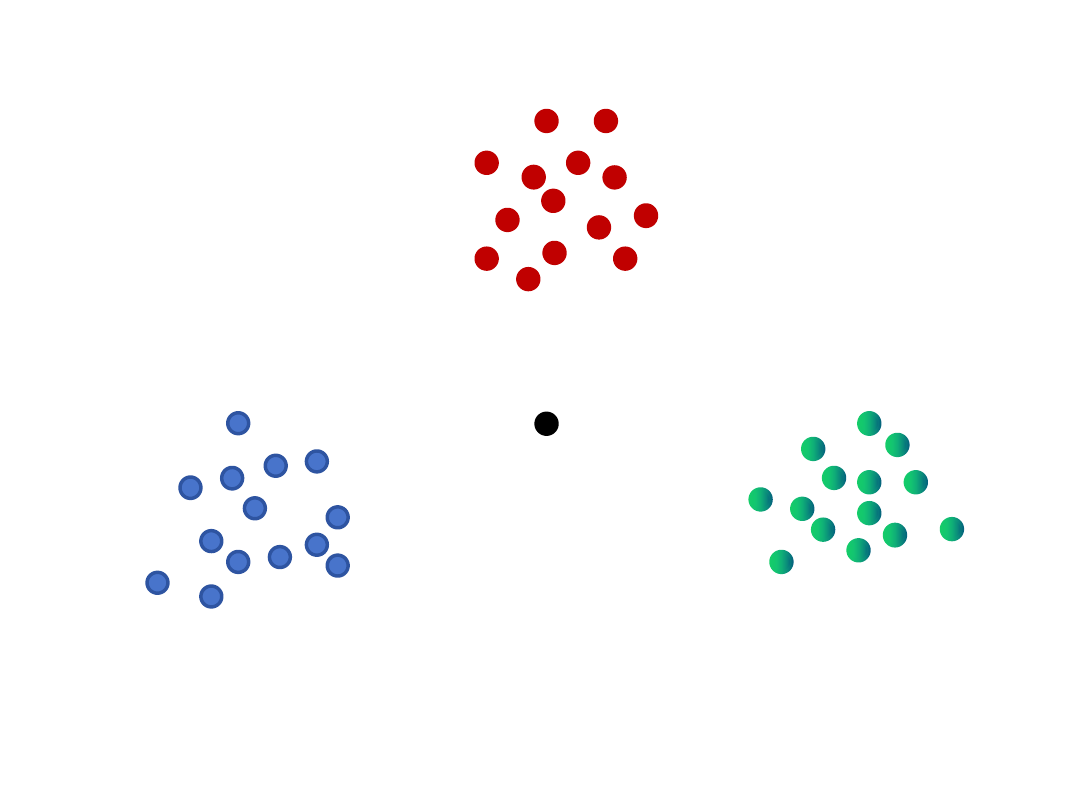}
\caption{\label{fig:kmeans} The K-means algorithm can divide the background events~(the colored points) into several groups. The anomalous events are expected to be far away from all centroids of the groups. An example of the anomaly is depicted as the black point.}}
\end{figure}

Another algorithm along similar lines to IF is an anomaly detection algorithm based on K-means.
K-means-based anomaly detection~(KMAD) is a method of anomaly detection that uses the K-means clustering algorithm to identify anomalous points in a data set.

The K-means algorithm is a popular clustering algorithm that partitions a data set into K clusters, where each point belongs to the cluster with the closest centroid~\cite{kmeans}. 
In KMAD, the K-means algorithm is used to cluster the normal points in the data set, and the distance from a point to the nearest centroid of the clusters is used to quantify the degree of anomaly of the point~\cite{Zhang:2023yfg}.

The basic idea of KMAD is that anomalous points are those that are far from all of the centroids of the normal clusters. 
The distance of a point from the nearest centroid can be used as a measure of its degree of anomaly. 
That is, the background is categorized using the K-means algorithm, such that the distribution of the background events can be described and sampled by a set of centroids around which the events are distributed. 
If the anomalous events are kinematically different from the ones of the background, the anomalous events will be farther away from all the centroids. 
Therefore, if a point is farther away from all the centroids than a certain threshold, it can be considered an anomaly.
An illustration of KMAD is shown in Fig.~\ref{fig:kmeans}.

In Ref.~\cite{Zhang:2023yfg}, the KMAD is used to study the NP contribution in the process $\mu^+\mu^-\to \gamma\gamma \gamma$, which is an example that the KMAD works well.

%%%%%%%%%%%%%%%%%%%%%%%%%%%%%%%%%%%%%
\section{\label{sec4}Typical usage}
%%%%%%%%%%%%%%%%%%%%%%%%%%%%%%%%%%%%%

To summarize, data preparation and feature engineering are critical steps in ML that cannot be overlooked. 
The quality and relevance of the features can have a significant impact on the performance of the model. 
Therefore, it is important to spend sufficient time and effort in preparing the data and engineering the features to ensure the best possible outcome.

In this section, we present a few examples on how to use the \verb"MLAnalysis" in the data preparation phase, as well as how \verb"MLAnalysis" can be used directly to select event signals.

\subsection{\label{sec4.1} Build a data set for KMAD, IF and NIF algorithm}

Data preparation and data culling are necessary aspects before ML algorithms. 
In the case of the process $\mu^+\mu^-\to \gamma\gamma\gamma$, for example, the final state sometimes contains less than two photons due to detector simulations. 
At this point, frequently, retaining only events containing three final-state photons improves the accuracy of signal screening. 
For the same reason, the final state sometimes contains more than three photons, at which point we can select the hardest three photons and compose their four-momenta into a 12-dimensional vector for each event. 
The 12-dimensional vectors are then to be stored in a ``csv'' file for use in the next KMAD, IF or NIF algorithms. 
The following code is an example of using \verb"MLAnalysis" to accomplish the above.

\begin{lstlisting}
def ChooseEventWithStrategy(allEvents: EventSet, count: int, tag: int):
    result = []
    idx = 0
    while len(result) < count:
        theEvent = allEvents.events[idx]
        largestPhotonIndex = -1
        largestPhotonEnergy = 0
        secondPhotonIndex = -1
        secondPhotonEnergy = 0
        thirdPhotonIndex = -1
        thirdPhotonEnergy = 0
        for theParticle in theEvent.particles:
            if theParticle.particleType == ParticleType.Photon:
                PhotonEnergy = theParticle.momentum.Momentum()
                if PhotonEnergy > largestPhotonEnergy:
                    thirdPhotonIndex = secondPhotonIndex
                    thirdPhotonEnergy = secondPhotonEnergy
                    secondPhotonIndex = largestPhotonIndex
                    secondPhotonEnergy = largestPhotonEnergy
                    largestPhotonIndex = theParticle.index - 1
                    largestPhotonEnergy = PhotonEnergy
                elif PhotonEnergy > secondPhotonEnergy:
                    thirdPhotonIndex = secondPhotonIndex
                    thirdPhotonEnergy = secondPhotonEnergy
                    secondPhotonIndex = theParticle.index - 1
                    secondPhotonEnergy = PhotonEnergy
                elif PhotonEnergy > thirdPhotonEnergy:
                    thirdPhotonIndex = theParticle.index - 1
                    thirdPhotonEnergy = PhotonEnergy
        if largestPhotonIndex >= 0 and secondPhotonIndex >= 0 and thirdPhotonIndex >= 0:
            toAdd = theEvent.particles[largestPhotonIndex].momentum.values
            toAdd = toAdd + theEvent.particles[secondPhotonIndex].momentum.values
            toAdd = toAdd + theEvent.particles[thirdPhotonIndex].momentum.values
            toAdd = toAdd + [tag]
            result.append(toAdd)
        idx = idx + 1
    return result
\end{lstlisting}

The above code loads data from an LHCO file and converts the data to the format which is ready for the KMAD algorithm. 
The data set contains information about particle collisions involving the production of three photons.

The main function in the first code, ``ChooseEventWithStrategy'', works by iterating over each event in ``allEvents'', selecting the three photons with the largest energies, and combining their momenta into a single vector along with the tag. 
It keeps track of the largest, second-largest, and third-largest photons, and updates these values as it iterates over the particles in the event. 
If it finds an event with at least three photons, it adds the combined momentum vector to the result list.
If result contains ``count'' elements, the function returns the result list.

\begin{lstlisting}
import os

from Applications.kmeans.kmeansfunctions import ChooseEventWithStrategy, SaveCSVFile
from CutAndExport.CutEvent import CutEvents
from CutAndExport.CutFunctions import PhotonNumberCut
from Interfaces.LHCOlympics import LoadLHCOlympics

os.chdir("../../")


headList = ["FT0", "FT2", "FT5", "FT7", "FT8", "FT9"]
energyList = ["1500", "5000", "7000", "15000"]
PhotonNumberCut = PhotonNumberCut(1, [3])

for he in headList:
    for en in energyList:
        for i in range(0, 11):
            testEvent = LoadLHCOlympics("_DataFolder/triphoton/cs/{0}/{0}-{1}-{2}.lhco".format(he, en, i))
            CutEvents(testEvent, PhotonNumberCut)
            resultList = ChooseEventWithStrategy(testEvent, len(testEvent.events), 0)
            toSave = "_DataFolder/kmeans/cs/E{0}/{1}/{1}-{0}-{2}.csv".format(en, he, i)
            SaveCSVFile(toSave, resultList)
            print(toSave, " saved! with events: ", len(testEvent.events))
\end{lstlisting}

The above code imports various functions and libraries, including ``os'' for directory navigation and ``LoadLHCOlympics'' for loading the LHCO data set. 
It sets up some variables: ``headList'', ``energyList'', and ``PhotonNumberCut'', in which ``headList'' and ``energyList'' are lists of strings representing different types of events in the LHCO files, while ``PhotonNumberCut'' is a Cut object used to filter events with fewer than 3 photons introduced in the previous section.
The ``PhotonNumberCut'' is applied as the data cleaning phase.
The strings in ``headList'' correspond to the origins of the NP signals, i.e. the $O_{T_i}$ operators contributing to the aQGCs.
The strings in  ``energyList'' correspond to the beam energies of the muon colliders.
The code loops over each combination of ``headList'', ``energyList'', and event index ``i''. 
For each combination, it loads the corresponding LHCO file using ``LoadLHCOlympics'', applies the ``PhotonNumberCut'' to filter out events with fewer than 3 photons using ``CutEvents'', chooses events with the ``ChooseEventWithStrategy'' function, saves the results to a ``csv'' file using ``SaveCSVFile''. 
Finally, the code prints the file name and the number of events saved.

After the ``csv'' files are saved, they can be directly used in the KMAD, IF and NIF algorithms to investigate the efficiency of the algorithms in searching for NP signals.

In the case of KMAD, \verb"MLAnalysis" have functions ``KMeans'' and ``CalculateDistance'' in ``Applications/kmeans/kmeansfunctions'',

\verb!def KMeans(dataList, d: int, k: int, nmin: int = 0) -> bool:!

The ``KMeans'' function is fed with a data table as ``dataList'', and number of features are provided as ``d'', the number of clusters are provided as ``k''.
The data table is required to have one more column than the number of features which is used to store the cluster assignments of each point.
There is an optional parameter ``nmin'', which specifies a number of points~(denoted as $n_{min}$).
When the number of points which change the cluster assignments is smaller than $n_{min}$, the function will stop.
The function fails when there is at least a cluster without events.
The returned value specifies whether the function succeeds.
The cluster assignments are stored as the last elements of the vectors.

\verb!def CalculateDistance(dataList, d: int, k: int):!

The input parameters of function ``CalculateDistance'' are the same as the ``KMeans'' function.
The data table is required to have one more column than the number of features which stores the cluster assignments of the points.
The returned value is an array storing the minimum distances for all points, where the minimum distance for a point is defined as the minimal distance between the point to all centroids. 

The following code is an example to use the KMAD.

\begin{lstlisting}
import numpy as np
from Applications.kmeans.kmeansfunctions import KMeans, CalculateDistance
    
k = 50
l = 100
dim = 12
data = np.loadtxt(data.csv", delimiter=',')
averageDistance = []
for i in range(0, l):
    succeed = False
    while not succeed:
        succeed = KMeans(data, dim, k)
    np.savetxt("cluster-{0}.csv".format(i), data[:, dim].astype(int), delimiter=',', fmt='%i')
    distance = CalculateDistance(data, dim, k)
    averageDistance.append(distance)
npAllDistance = np.array(averageDistance)
npAllDistance = np.transpose(npAllDistance)
npAverageDistance = np.mean(npAllDistance, axis=1)   
\end{lstlisting}    

The above code loads a data table from a ``csv'' file, and then use the ``KMeans'' function to apply the K-means algorithm.
Then, the cluster assignments are stored for further usages.
The minimum distances are calculated using ``CalculateDistance''.
Since the result depends on the randomly initialed cluster assignments, the code repeatedly calculates minimum distances for $l=100$ times and stores the results in ``npAllDistance''.
The average minimum distance ``npAverageDistance'' is then calculated as the anomaly scores.

The core of both the IF and NIF algorithm is the construction of an iTree.
\verb"MLAnalysis" has a function called ``Split'' in ``Applications/IsolationForest/IsolationTree.py'' to construct an iTree.

\verb!def Split(dataArray, length: int, maxSplit: int):!

The ``Split'' function built an iTree with the data fed as ``dataArray'', and number of features are provided as ``length''.
Two columns need to be added at the end of the data table to be used as results. 
The first of these columns is used to hold the source label of the data (e.g., whether it is from background or signal), and the function ``Split'' will not use this information or change the results in this column. 
The second column is used to hold the depth of the leaf in the iTree corresponding to a piece of data.
The parameter ``maxSplit'' specifies the maximum depth of the leafs of an iTree.
When ``maxSplit'' is set to $-1$, the iTree will be built with one point on one leaf.
The returned values are the data sets with the last column set as the depths of the leafs.

The following code is an example to use the ``Split'' function to build iTrees.

\begin{lstlisting}
Loop = 100
L = 18
saveCol = [L, L + 1]
dataSet1 = np.loadtxt("background.csv", delimiter=',')
dataSet2 = np.loadtxt("signal.csv", delimiter=',')
length = len(dataSet1)
z1 = np.zeros([length, 2])
length = len(dataSet2)
o2 = np.ones([length, 1])
z2 = np.zeros([length, 1])
dataSet1 = np.hstack((dataSet1, z1))
dataSet2 = np.hstack((dataSet2, o2, z2))
dataSet = np.vstack((dataSet1, dataSet2))
for i in range(0, Loop):
    print("======== loop {} ==========".format(i + 1))
    resSet = Split(dataSet, L, -1)
    np.savetxt(saveName + str(i) + ".csv", resSet[:, saveCol].astype(int), delimiter=',', fmt='%i')
\end{lstlisting}  

The above code loads two data tables with $L=18$ features, and then use the ``Split'' function to build $Loop=100$ iTrees~(i.e. an IF).
Then, the tags and the depths of the leafs are stored in a ``csv'' file for further usages.
For example, the average depths can be directly used as the anomaly scores in the IF algorithm.

\subsection{\label{sec4.2} Using artificial neural network to reconstruct the energy of subprocess}

For an effective field theory~(EFT), the Wilson coefficients are usually functions of energy scales. 
Apart from that, the energy scale of the process is the basis for studying the validity of the EFT.
However, the energy of the VBS subprocess can only be reconstructed by the final state particles. 
If there are multiple neutrinos in the final state, the traditional method based on kinematics is difficult to reconstruct the energy of the process.
Artificial neural networks (ANN) are effective at finding complex relationships between inputs and outputs, which is suitable for solving such problems.
Therefore, in the following example, we will present the data preparation phase of using the ANN to reconstruct the energy scale of a VBS subprocess with multiple neutrinos in the final state which is used in Ref.~\cite{Yang:2021ukg}.

Before using ANN, we need training data sets and validation data sets. The LHE files created by \verb"MadGraph5_aMC@NLO" contain the information of neutrinos, so the $\hat{s}$ can be calculated. 
With $\hat{s}$ calculated, it is possible to build the label for the output layer of the ANN.

\begin{lstlisting}
def SHatWWReal(eventSample: EventSample) -> float:
    """
    pW + pW = pl + pl + pnu + pnu
    assume LHE file, so nu is visible
    :param eventSample:
    :return:
    """
    pall = LorentzVector(0, 0, 0, 0)
    for i in range(len(eventSample.particles)):
        if 1 <= eventSample.particles[i].particleType <= 2:
            pall = pall + eventSample.particles[i].momentum
        if ParticleType.Missing == eventSample.particles[i].particleType:
            pall = pall + eventSample.particles[i].momentum
    return pall * pall
\end{lstlisting}

This code defines a function called ``SHatWWReal'' that takes an object of type ``EventSample'' as input and returns a float value. 
The function assumes that the input EventSample object is from an LHE file so that the neutrinos are visible.

The function calculates the square of the total momentum of the system resulting from the collision of two $W$ bosons, which decay into two charged leptons and two neutrinos.
The ``pall'' variable is initialized as a Lorentz vector with zero momentum and iterates over all the particles in the given event sample. 
If the particle type is 1 or 2, which represent the charged leptons~(Electron=1, Muon=2) , then their momenta are added to the total four-momentum. 
If the particle type is ``Missing'', which represents the neutrinos, then their momenta are also added to the total four-momentum. 
Finally, the function returns the square of the total momentum of the system.

The output of this function is used as a label for the output layer of an ANN that is trained to reconstruct the energy of a sub-process in a collision event.
When `SHatWWReal' function is implemented, we need to read an LHCO file and the corresponding LHE file to build the training and validation data sets.

\begin{lstlisting}
def Export(eventSetLHCO, eventSetLHE, startIndex, endIndex, applyCut, file):
    normalizer = 1000.0
    # result_f.write("j1x,j1y,j1z,j2x,j2y,j2z,l1x,l1y,l1z,l2x,l2y,l2z,mx,my,shat\n")
    for i in range(startIndex, endIndex):
        oneEvent = eventSetLHCO.events[i]
        lepton1 = LorentzVector(0, 0, 0, 0)
        lepton2 = LorentzVector(0, 0, 0, 0)
        jet1 = LorentzVector(0, 0, 0, 0)
        jet2 = LorentzVector(0, 0, 0, 0)
        missing = LorentzVector(0, 0, 0, 0)
        largestJetIndex1 = 0
        largestJetM1 = 0.0
        largestJetIndex2 = 0
        largestJetM2 = 0.0
        leptonIdx1 = 0
        leptonIdx2 = 0
        largestLepton1 = 0
        largestLepton2 = 0
        hasMissing = False
        for oneParticle in oneEvent.particles:
            if ParticleStatus.Outgoing == oneParticle.status \
                    and ParticleType.Jet == oneParticle.particleType:
                momentum = oneParticle.momentum.Momentum()
                if momentum > largestJetM1:
                    largestJetM2 = largestJetM1
                    largestJetIndex2 = largestJetIndex1
                    largestJetM1 = momentum
                    largestJetIndex1 = oneParticle.index
                elif momentum > largestJetM2:
                    largestJetM2 = momentum
                    largestJetIndex2 = oneParticle.index
            elif ParticleType.Electron <= oneParticle.particleType <= ParticleType.Muon:
                momentumLepton = oneParticle.momentum.Momentum()
                if oneParticle.PGDid > 0 and momentumLepton > largestLepton1:
                    leptonIdx1 = oneParticle.index
                    largestLepton1 = momentumLepton
                elif oneParticle.PGDid < 0 and momentumLepton > largestLepton2:
                    leptonIdx2 = oneParticle.index
                    largestLepton2 = momentumLepton
            elif ParticleType.Missing == oneParticle.particleType:
                hasMissing = True
                missing = missing + oneParticle.momentum
        if not (leptonIdx1 > 0 and leptonIdx2 > 0):
            continue
        if not (largestJetIndex1 > 0 and largestJetIndex2 > 0):
            continue
        if not hasMissing:
            print(oneEvent.DebugPrint())
            continue
        lepton1 = oneEvent.particles[leptonIdx1 - 1].momentum
        lepton2 = oneEvent.particles[leptonIdx2 - 1].momentum
        jet1 = oneEvent.particles[largestJetIndex1 - 1].momentum
        jet2 = oneEvent.particles[largestJetIndex2 - 1].momentum
        if applyCut:
            if lepton1.values[1] * lepton1.values[1] + lepton1.values[2] * lepton1.values[2] < 1.0e2:
                continue
            if lepton2.values[1] * lepton2.values[1] + lepton2.values[2] * lepton2.values[2] < 1.0e2:
                continue
            if missing.values[1] * missing.values[1] + missing.values[2] * missing.values[2] < 1.0e2:
                continue
            lepX = lepton1.values[1] + lepton2.values[1]
            lepY = lepton1.values[2] + lepton2.values[2]
            if lepX * lepX + lepY * lepY < 100:
                continue
            lengthLep = math.sqrt(lepX * lepX + lepY * lepY)
            lengthM = math.sqrt(missing.values[1] * missing.values[1] + missing.values[2] * missing.values[2])
            dotLM = (lepX * missing.values[1] + lepY * missing.values[2]) / (lengthLep * lengthM)
            if abs(dotLM) < 0.8:
                continue
            lengthL1 = math.sqrt(lepton1.values[1] * lepton1.values[1]
                                 + lepton1.values[2] * lepton1.values[2]
                                 + lepton1.values[3] * lepton1.values[3])
            lengthL2 = math.sqrt(lepton2.values[1] * lepton2.values[1]
                                 + lepton2.values[2] * lepton2.values[2]
                                 + lepton2.values[3] * lepton2.values[3])
            dotLL = (lepton1.values[1] * lepton2.values[1]
                     + lepton1.values[2] * lepton2.values[2]
                     + lepton1.values[3] * lepton2.values[3]) / (lengthL1 * lengthL2)
            if dotLL > -0.5:
                continue
        realShat = SHatWWReal(eventSetLHE.events[i])
        paramLst = [jet1.values[0] / normalizer,
                    jet1.values[1] / normalizer,
                    jet1.values[2] / normalizer,
                    jet1.values[3] / normalizer,
                    jet2.values[0] / normalizer,
                    jet2.values[1] / normalizer,
                    jet2.values[2] / normalizer,
                    jet2.values[3] / normalizer,
                    lepton1.values[0] / normalizer,
                    lepton1.values[1] / normalizer,
                    lepton1.values[2] / normalizer,
                    lepton1.values[3] / normalizer,
                    lepton2.values[0] / normalizer,
                    lepton2.values[1] / normalizer,
                    lepton2.values[2] / normalizer,
                    lepton2.values[3] / normalizer,
                    missing.values[1] / normalizer,
                    missing.values[2] / normalizer]
        strW = ""
        for x in range(0, 18):
            strW = "{}{:.5e},".format(strW, paramLst[x])
        strW = "{}{:.5e}\n".format(strW, math.sqrt(realShat) / normalizer)
        file.write(strW)
\end{lstlisting}

The code above defines a function called ``Export'' that takes in six parameters: ``eventSetLHCO'', ``eventSetLHE'', ``startIndex'', ``endIndex'', ``applyCut'', and ``file''. Inside the function, a variable normalizer is initialized with a value of $1000.0$, the goal is to change the unit from GeV to TeV.

The data in the above code is arranged as a $19\times N$ table, where $N$ is the number of events, each line is a 19 dimensional vector represents an event. 
They are the components of the 4-momenta of the two hardest jets, the 4-momenta of the two hardest opposite signed charged leptons and the components of the transverse missing momentum, they are all observables. The output of the ANN corresponds to $\hat{s}$. The true labels are the 19-th variables of the elements in the data sets which are $\hat{s_{\rm tr}}$ of the events.

At the beginning of the code, we iterate through each particle in the event, selecting the two jets with the highest momenta and the two leptons (either electrons or muons) with the highest momenta. We require that these two leptons are a pair of oppositely charged leptons. 
Additionally, the function checks if there is any missing momentum in the event.
It can be found that not all events are useful, such as the events where two jets are not found, events where two leptons with the largest momenta are not a pair of opposite-charged particles, and events where missing energy is not found. 
Therefore, to ensure the effectiveness of the subsequent code, we perform data cleaning on events.

In this study, the NP is introduced via the SMEFT.
As an EFT, the $\hat{s}$ is important.
However, we do not need to reconstruct the $\hat{s}$ of the SM events.
To increase the efficiency of the ANN, we can remove the SM events. 
We introduce a switch called ``applyCut'' in the second part of the code, which includes a series of filtering strategies: requiring the transverse momentum (``pT'') of each charged lepton to be greater than 100 GeV, requiring the missing transverse momentum to be greater than 100 GeV, requiring the transverse momentum of the lepton pair combination to be greater than 100 GeV, requiring the cosine of the angle between the transverse momentum of the lepton pair and the missing transverse momentum to be greater than 0.8, and requiring the cosine of the angle between the pair of oppositely charged leptons to be less than -0.5.
If an event passes all the selection criteria, it is more likely from the NP and is written to the output file. 
Those cuts are the cuts to high light the NP contribution used in Ref.~\cite{Guo:2019agy}.
When ``applyCut'' is set to true, the occurrence of background events can be suppressed, resulting in training and validation data sets that mainly contain signal events.

The end part of the code writes the selected events and their corresponding input parameters to a file.
Afterwards, the parameters are appended to the string ``strW'' using Python's ``format'' function and written to the file using Python's ``write'' function. 
Each line of the file corresponds to an event, with the input parameters and output values separated by commas.

Finally, the data is saved as a ``csv'' file, and is ready to be fed to the ANN~\cite{Yang:2021ukg,tensorflow}.

\subsection{\label{sec4.3} ESS using polarization}

In order to distinguish between the SM events and NP events, one can utilize certain special properties such as polarization. 
Take the VBS process as an example, for many events, the charged leptons in the final states are from the $Z$ boson decays, whose polarizations can be transverse or longitudinal.
In the rest frame of a $Z$ boson with the flight direction of the $Z$ boson as the ${\bf z}$-axis, the charged leptons in the $Z$ boson decay satisfy an angular distribution~\cite{Peruzzi:2011mqa}, 
\begin{equation}
\frac{d\sigma}{d\cos \theta^*}\propto f_{L}\frac{\left(1-\cos \theta^*\right)^2}{4} + f_R \frac{\left(1+\cos \theta^*\right)^2}{4} + f_0\frac{\sin ^2 \theta^*}{2},
\label{eq.polarization}
\end{equation}
where $f_{L,R,0}$ represent the fraction of left, right-handed and longitudinal polarizations, and $\theta^*$ is the zenith angle of the charged lepton in the rest frame of $Z$ boson.

In the SM, $Z$ bosons are mainly longitudinally polarized. 
In some NP models, $Z$ bosons are mainly transversely polarized, leading to a different angular distribution. 
Therefore, we can extract NP signal events from the SM background by utilizing the angular distribution of the charged leptons produced by the decay of the $Z$ boson in its rest frame.

In the code below, we define a function called ``leptonPZ''. 
The first 23 lines of the function traverse all particles in the event and select the two leptons with the largest momenta (``largestLM1'' and ``largestLM2''). 
Then, the sum of the momenta of the two selected leptons is assigned to ``pZ'', and the momentum of the positive charged lepton is assigned to ``pL''. 
That is, we reconstruct the momentum of the $Z$ boson (``pZ'') from the sum of the momenta of the lepton pair produced by its decay.

In the latter part of the code, we use two functions in ``Matrix4x4''. 
First, we use the ``MakeRotation'' function to create a rotation matrix~(``rotMtr'') that rotates the momentum of the $Z$ boson (``pZ'') such that it aligns with the ${\bf z}$-axis. 
It then applies the same rotation to the momentum of the lepton~(``pL'').
This rotation matrix acts as a rotation of the frame, and ``pLDir'' is the result of the momentum of the charged lepton after the rotation is applied, where the rotation is the one rotates the momentum of the $Z$ boson to the ${\bf z}$-axis.
Then, we use the ``MakeBoost'' function to generate a Lorentz transformation matrix that transforms ``pLDir'' into the rest frame of the $Z$ boson, obtaining ``pLRest''. Finally, the function returns the cosine of the zenith angle of ``pLRest'' in the rest frame of the $Z$ boson~(the $\cos (\theta^*)$ in Eq.~(\ref{eq.polarization})).
In this way, one can apply a cut on $\cos (\theta ^*)$ to highlight the NP signals if the polarizations of the gauge bosons are different from the SM~\cite{Guo:2020lim,Yang:2021pcf,Fu:2021mub}.

\begin{lstlisting}
def LeptonPZ(eventSample: EventSample) -> float:
    largestLM1 = 0
    largestLM2 = 0
    largestLIndex1 = 0
    largestLIndex2 = 0
    for particle in eventSample.particles:
        if ParticleType.Electron == particle.particleType or ParticleType.Muon == particle.particleType:
            momentum = particle.momentum.Momentum()
            if momentum > largestLM1:
                largestLM2 = largestLM1
                largestLIndex2 = largestLIndex1
                largestLM1 = momentum
                largestLIndex1 = particle.index
            elif momentum > largestLM2:
                largestLM2 = momentum
                largestLIndex2 = particle.index
    p41 = eventSample.particles[largestLIndex1 - 1].momentum
    p42 = eventSample.particles[largestLIndex2 - 1].momentum
    pZ = p41 + p42
    pL = p41
    if eventSample.particles[largestLIndex2 - 1].PGDid > 0:
        pL = p42
    rotMtr = Matrix4x4.MakeRotationFromTo(pZ.V3d(), [0, 0, 1])
    pZDir = rotMtr.MultiplyVector(pZ)
    pLDir = rotMtr.MultiplyVector(pL)
    v3dz = pZDir.V3d()
    vsq = Constants.dot3d(v3dz, v3dz)
    if vsq > 0.9999999999 or vsq < 0:
        # This event sample should be excluded
        return 1000
    boostMtr = Matrix4x4.MakeBoost(pZDir.V3d())
    pLRest = boostMtr.MultiplyVector(pLDir)
    return math.cos(pLRest.Theta())
\end{lstlisting}

%%%%%%%%%%%%%%%%%%%%%%%%%%%%%%%%%%
\section{\label{sec5}Summary}
%%%%%%%%%%%%%%%%%%%%%%%%%%%%%%%%%%

ML algorithms are widely used in many fields of HEP.
\verb!MLAnalysis! provides a suite of tools for data transformation and feature engineering. 
It converts experimental or simulated data into a data set that can be used for ML, through which a variety of algorithms such as ANN can be easily applied. 
The current version of the \verb!MLAnalysis! has built-in ML algorithms including IF, NIF and KMAD.  
These ML-based approaches can help researchers to optimize ESS to improve signal significance. 

In Sec.~\ref{sec3}, we detail the IF, NIF and KMAD algorithms and their applications. 
IF and KMAD algorithms are AD algorithms which are NP model-independent and SMEFT operator-independent.
No matter what the NP signals exist in the data set, they can be detected as long as they satisfy ``few and different'' from the SM events.
NIF algorithm not only inherits the advantages of IF algorithm, but also can deal with non-AD problems, such as detecting NP effects dominated by interference.
As automatic ESSs, IF, NIF and KMAD can achieve signal recognition ability comparable to or even better than that of a traditional ESS without kinematic analysis, and greatly improve the analysis efficiency without knowing what kind of NP model or SMEFT operator one is studying.
We presented several examples of \verb!MLAnalysis! in Sec.~\ref{sec4}, which include how to build a data set for ML, how to reconstruct the energy scale of the subprocess and how to optimize the ESS using polarization. 

The combination of data preparation and ESSs in \verb!MLAnalysis! makes it a comprehensive and valuable tool for researchers working with experimental and MC simulation data. 
It has the potential to make ML more accessible and effective in the NP research. 
If users want to apply ML algorithms to other problems beyond the scope of this paper, the open-source \verb!MLAnalysis! based on user-friendly Python code enables a path forward for exploration.

\section*{Conflict of interest}
The authors declare that they have no known competing financial interests or personal relationships that could have appeared to influence the work reported in this paper.

\section*{Acknowledgment}

\noindent
This work was supported in part by the National Natural Science Foundation of China under Grants Nos. 11905093 and 12147214, the Natural Science Foundation of the Liaoning Scientific Committee No.~LJKZ0978 and the Outstanding Research Cultivation Program of Liaoning Normal University (No.21GDL004) and ``New strategies for detecting signal of new physics at the future lepton colliders'', and the University-Industry Collaborative Education Program No.~220800575313412.

%\appendix
%\section{\label{ap1}The class of program}

\begin{comment}
\begin{table}
  \centering
    \caption{ParticleTable}\label{tab:6}
  \begin{tabular}{|l|}
     \hline
\multicolumn{1}{|c|}{ParticleTable}\\
     \hline
EWcharge\\
PGDid\\
mass\\
spin\\
width\\
     \hline
   \end{tabular}
\end{table}
\end{comment}

%\bibliography{MLAnalysis}
%\bibliographystyle{elsarticle-num}

\end{document}